

\input phyzzx


\def\Emissperp{\not \! \!  E_{T}}
\def\EmissT{\not \! \!  E_{T}}

\def\Emiss{\not  \! \! E}

\def\ls#1{_{\lower1.5pt\hbox{$\scriptstyle #1$}}}

\frontpagetrue


\let\picnaturalsize=N
\def\picsize{1.0in}
\def\picfilename{scipp_tree.eps}

\let\nopictures=Y

\ifx\nopictures Y\else{\ifx\epsfloaded Y\else\input epsf \fi
\let\epsfloaded=Y
{\line{\hbox{\ifx\picnaturalsize N\epsfxsize \picsize\fi 
{\epsfbox{\picfilename}}}\hfill\vbox{


\hbox{SCIPP 96-08} 
\hbox{SU-ITP-96-06}
\hbox{SLAC-PUB-96-7104} 
\hbox{hep-ph/9601367}
\vskip1.2in
}
}}}\fi


\overfullrule 0pt

\pubtype{ T}     

\rightline{SLAC-PUB-96-7104} \vskip-8pt
\rightline{SU-ITP-96-06} \vskip-8pt
\rightline{SCIPP 96-08} \vskip-8pt
\rightline{hep-ph/9601367} \vskip-8pt

\title{\seventeenbf Experimental Signatures of Low Energy }
\vskip-.4in

\title{\seventeenbf Gauge Mediated Supersymmetry Breaking}
\author{Savas Dimopoulos$^{ab}$, Michael Dine$^c$\foot{Work 
supported by the Department of Energy.}, 
Stuart Raby$^d$\foot{Work supported by the Department of 
Energy under contract DOE-ER-01545-646.}, 
Scott Thomas$^e$\foot{Work supported by the Department of 
Energy under contract DE-AC03-76SF00515.}}

\baselineskip=14pt
\address{$^a$Physics Department, Stanford University, Stanford, CA 94309}
\vskip-8pt
\address{$^b$Theoretical Physics Division, CH-1211, Geneva 23,
Switzerland}
\vskip-8pt
\address{$^c$Santa Cruz Institute for Particle Physics,
University of California, Santa Cruz, CA 95064}
\vskip-8pt
\address{$^d$Physics Department, Ohio State University,
Columbus, OH 43210}
\vskip-8pt
\address{$^e$Stanford Linear Accelerator Center,
Stanford, CA 94309}


\vskip.4in
\vbox{
\centerline{\bf Abstract}


The experimental signatures for low energy gauge mediated supersymmetry
breaking are distinctive since the gravitino is naturally the LSP.
The next lightest supersymmetric particle (NLSP) can be a gaugino,
Higgsino, or right handed slepton.
For a significant range of parameters decay of the NLSP to its
partner plus the gravitino can be measured as a displaced vertex
or kink in a charged particle track. 
In the case that the NLSP is mostly gaugino, we identify 
the discovery modes as $e^+e^- \rightarrow \gamma \gamma + \Emiss$,
and $p \bar{p} \rightarrow l^+ l^- \gamma \gamma + \EmissT$.
If the NLSP is a right handed slepton the discovery modes are 
$e^+ e^- \rightarrow l^+ l^- + \Emiss$ and 
$p \bar{p} \rightarrow l^+ l^- + \EmissT$.
An NLSP which is mostly Higgsino is also considered.




}


\parskip 0pt
\parindent 25pt
\overfullrule=0pt
\baselineskip=19pt
\tolerance 3500
\endpage
\pagenumber=1
\singlespace
\bigskip


\singlespace
\bigskip

\chapter{Introduction}

\baselineskip=18pt




\REF\lsgauge{M. Dine, W. Fischler, and M. Srednicki, Nucl. Phys. B 
{\bf 189} (1981) 575;
S. Dimopoulos and S. Raby, Nucl. Phys. B {\bf 192} (1981) 353;
M. Dine and W. Fischler, Phys. Lett. B {\bf 110} (1982) 227;
M. Dine and M. Srednicki, Nucl. Phys. B {\bf 202} (1982) 238;
L. Alvarez-Gaum\'{e}, M. Claudson, and M. Wise, Nucl. Phys. B 
{\bf 207} (1982) 96;
C. Nappi and B. Ovrut, Phys. Lett. B {\bf 113} (1982) 175.}

\REF\hsgauge{M. Dine and W. Fischler, Nucl. Phys. B {\bf 204} (1982) 346;
S. Dimopoulos and S. Raby, Nucl. Phys. B {\bf 219} (1983) 479.}


\REF\dnns{M. Dine, A.E. Nelson and Y. Shirman, 
Phys. Rev. D {\bf 51} (1995) 1362;
M. Dine, A.E. Nelson, Y. Nir and Y. Shirman,
Phys. Rev. D {\bf 53} (1996) 2658.} 

\REF\soft{S. Dimopoulos and H. Georgi, Nucl. Phys. B {\bf 193} (1981) 150.}

\REF\dimsut{S. Dimopoulos and D. Sutter, Nucl. Phys. B
{\bf 452} (1995) 496. 
}
\REF\il{I.E. Ibanez and D. Lust, Nucl. Phys. B
{\bf 382} (1992) 305.}

\REF\kl{V. Kaplunovsky and J. Louis, Phys. Lett. B
{\bf 306} (1993) 269.}

\REF\niretal{M. Leurer, Y. Nir and N. Seiberg,
Phys. Lett. B {\bf 309} (1993) 337. 
}
\REF\dkl{M. Dine, A. Kagan and R. Leigh,
Phys. Rev. D {\bf 48} (1993) 4269. 
}
\REF\kaplan{D.B. Kaplan and M. Schmalz,
Phys. Rev. D {\bf 49} (1994) 3741. 
}

Low energy supersymmetry is widely viewed as a plausible
solution of the hierarchy problem.  If nature is supersymmetric,
it is important to understand how supersymmetry is broken.
It is usually assumed that supersymmetry breaking is
communicated to ordinary fields and their
superpartners by supergravity.   The breaking
scale is then necessarily of order $10^{11}$ GeV.  
An alternative possibility,
which has been less thoroughly explored, is that supersymmetry
is broken at some lower energy scale, and that 
the ordinary gauge interactions
act as the messengers of supersymmetry 
breaking\refmark{\lsgauge-\dnns}.  In this
case, the scale of supersymmetry breaking can be as low as
$10$'s of TeV\refmark{\lsgauge,\dnns}.

Independent of source and messenger, supersymmetry
breaking is represented among ordinary fields (the
visible sector) by soft supersymmetry breaking terms
\refmark{\soft}.
 The most general soft-breaking Lagrangian is described
by $105$ parameters beyond those of the minimal
standard model\refmark{\dimsut}.  
There are a number of constraints which these 
parameters must satisfy, coming from direct experimental
searches for superpartners, electric dipole moments,
and the lack of flavor changing neutral currents. 
Most model builders simply postulate a high degree of degeneracy among
squarks and sleptons at a high energy scale to deal with this 
problem\refmark{\soft}.  
In certain classes of 
superstring theories, there are weak hints for 
such a universality\refmark{\il,\kl}.  Alternatively,
the various experimental constraints might
be satisfied as a result of flavor symmetries
or by other means\refmark{\niretal-\kaplan}. 
With gauge mediated supersymmetry breaking
the entire soft breaking Lagrangian can
be calculated in terms of a small
number of parameters. In addition,
the regularities required 
to avoid flavor changing neutral currents
0are automatically obtained since the ordinary gauge
interactions do not distinguish generations. 
For these reasons, we believe the gauge-mediated
possibility should be taken seriously.

In this letter, we discuss some striking and distinctive
signatures of low energy gauge-mediated supersymmetry
breaking.  The first is the spectrum of superpartner
masses.  
These masses are functions of the gauge quantum numbers,
and are roughly in the ratio of the appropriate gauge couplings squared. 
In the simplest models, definite relations
exist among these masses. 
As a result, the lightest standard model superpartner is almost
inevitably either
a neutralino or a right-handed slepton.  
The second important
signature arises from the fact that the
lightest supersymmetric particle (LSP) is the gravitino. 
The lightest standard model superpartner is then the
next to lightest supersymmetric particle (NLSP).
Assuming
that $R$-parity is conserved, the principle decay of the NLSP
is then to its partner plus a gravitino.  The longitudinal
component of the gravitino -- the Goldstino -- couples
to matter with strength proportional to $F^{-1}$, where
$F$ is the scale of supersymmetry breaking.
For a plausible range of $F$, the decay length can be 
100's of $\mu$m  to meters. 
The decays can therefore take place within a detector. 
This leads to signatures for supersymmetry which are
distinct from the conventional minimal supersymmetric
standard model (MSSM), and with potentially
dramatic displaced vertices.

\chapter{Superpartner Spectrum}

\REF\kenscott{K. Intriligator and S. Thomas,
SLAC-PUB-95-7041, hep-th/9603158.}

In gauge mediated models, supersymmetry is broken in a messenger
sector which transforms under the standard model gauge group.
The matter fields in this sector are generally referred to as
messenger quarks and leptons. 
Supersymmetry breaking is transmitted to the visible sector
by ordinary gauge interactions.
To preserve the successful supersymmetric prediction of the low
energy $\sin^2 \theta_W$ it is sufficient that the messengers
form a GUT representation. 
In the simplest versions, the messenger fields are weakly
coupled, and possess the quantum
numbers of a single ${\bf 5} + \bar{\bf 5}$ of $SU(5)$, i.e. there
are triplets, $q$ and $\bar q$, and doublets $\ell$ and
$\bar \ell$.  They couple to a single gauge singlet field,
$S$, through a superpotential 
$W=\lambda_1 S q \bar q
+ \lambda_2 S \ell \bar \ell$.  
The field $S$ has
non-zero expectation values for both scalar and auxiliary
components, $ S$ and $ F_S $.
Integrating out the messenger sector then gives rise to 
gaugino masses at one loop.  
For $F_S \ll S $,
these masses are given by \refmark{\dnns}
$$
m_{\lambda_i}=c_i\ N {\alpha_i\over4\pi}\ \Lambda\ .
\eqn\gluinomass
$$
where $c_1 = {5 \over 3}, c_2=c_3=1$,
$\Lambda =  F_S / S $,
and 
for a more general messenger sector $N$ is the 
equivalent number of $SU(5)$ $\bf{5} + \bar{\bf 5}$ representations. 
The scalar masses squared arise at two-loops \refmark{\dnns}
$$
\tilde m^2 ={2 \Lambda^2} N
\left[
C_3\left({\alpha_3 \over 4 \pi}\right)^2
+C_2\left({\alpha_2\over 4 \pi}\right)^2
+{5 \over 3}{\left(Y\over2\right)^2}
\left({\alpha_1\over 4 \pi}\right)^2\right].
\eqn\squarks
$$
where $C_3 = {4 \over 3}$ 
for color triplets and zero for singlets, $C_2= {3 \over 4}$ for
weak doublets and zero for singlets, 
and $Y$ is the ordinary hypercharge normalized
as $Q = T_3 + {1 \over 2} Y$.
It should be stressed that $F_S$ is not necessarily
the intrinsic supersymmetry breaking scale, $F$, 
since the gauge singlet field may not be coupled directly 
to the supersymmetry breaking sector. 
For example, in the model of Ref. \refmark{\dnns}, 
$F \gg F_S$.  However, it is also perfectly possible that
$F \sim F_S$\refmark{\kenscott}.
While $F_S$ determines the superpartner masses, it is $F$ which 
determines the Goldstino coupling discussed in the next section.

These expressions for the masses possess a number of noteworthy features.
There is a hierarchy of masses, with colored particles
being the most massive, and $SU(3) \times SU(2)$ singlet
particles the lightest.
The gaugino masses are in the ratio $7:2:1$,
just as for supersymmetry breaking with universal
gaugino masses at a high scale. 
For $N=1$ the squark, left handed slepton, right handed slepton,
and bino (partner of the hypercharge
gauge boson) masses are in the ratio $11.6:~2.5~:~1.1~:1$.
In this case the bino is the natural candidate for the NLSP.
The gaugino masses grow as $N$, while the scalar masses
grow as $\sqrt{N}$.
For $N=2$ the above masses are in the ratio 
$10.6~:~2.3~:~1~:~1.3$.
In this case the right handed slepton is the candidate for the 
NLSP.

In more general models the above relations among the masses
can be modified. 
For example both 
\gluinomass\ and \squarks\ are corrected at 
${\cal O}(F/ \lambda S^2)$.  
Additional modifications can arise with several gauge singlet
fields coupling to $q \bar{q}$ and $\ell \bar{\ell}$.
In the model with
one singlet, the couplings $\lambda_1$ and $\lambda_2$
cancel out in the expressions for the masses, but this is not true
of the more general case.  As a result, both the ratios
of the squark and slepton masses and the ratio of these
masses to gaugino masses are modified.
More generally scalar masses require only supersymmetry
breaking, while gaugino masses require also that
$U(1)_R$ be broken to at most $R$-parity.
In principle $U(1)_R$ could effectively be broken at a lower scale
than supersymmetry, leading to gauginos which are much lighter
than the scalars. 

Perhaps a more interesting possibility is that the messenger
sector is strongly coupled.  
Gaugino masses can then arise directly from 
non-perturbative dynamics in the messenger sector,
$m_{\lambda} \sim \alpha \Lambda$.
The scalar masses require one perturbative gauge loop,
$\tilde{m}^2 \sim \alpha ( \alpha / 4 \pi) \Lambda^2$.
So in this case the gauginos are much heavier than the scalars,
and the natural candidate for the NLSP is the
right handed slepton. 
All of the possibilities given above for the messenger sector
have in common the feature that
masses for standard model superpartners
go roughly as gauge couplings squared, although
the relation of scalar to gaugino masses is model dependent.

The dimensionful terms 
which must arise in the Higgs sector 
$W= \mu H_1 H_2$, and $V=m_{12}^2 H_1 H_2 + h.c.$, 
do not follow directly from the anzatz of
gauge mediated supersymmetry breaking, and are model dependent. 
This is because these terms require that the Peccei-Quinn 
symmetry between $H_1$ and $H_2$ 
be broken by non-gauge interactions. 
Specific models with additional singlets and 
vector quarks have been constructed in which 
$\mu$ and $m_{12}^2$ do arise with reasonable magnitude \refmark{\dnns}.
Because the properties of the Higgs sector are not generic, 
we leave open the possibility that the lightest
electroweak neutralino is a general mixture of 
gaugino and Higgsino.

\chapter{Phenomenology}

\REF\fayet{P. Fayet, Phys. Lett. B {\bf 84} (1979) 416;
Phys. Rept. {\bf 105} (1984) 21.}

Perhaps the most dramatic consequence of low energy gauge
mediated supersymmetry breaking is that the gravitino
is the LSP.  
In the global limit the Goldstone fermion, or Goldstino,
of supersymmetry breaking is massless. 
In local supersymmetry, the Goldstino becomes the 
longitudinal component of the gravitino,
giving a gravitino mass (assuming the cosmological
constant vanishes) of 
$$
m_G = {F \over \sqrt{3} M_p} 
\simeq 2.5 ~\left( { F \over (100 ~{\rm TeV})^2 } \right) ~ {\rm eV}
\eqn\gravmass
$$
where $F$ is the supersymmetry breaking scale. 
The lightest standard model supersymmetric particle is
then the NLSP, and can 
decay to its partner and the gravitino.
The lowest order coupling of the Goldstino is 
fixed by the supersymmetric 
Goldberger-Treiman low energy theorem to 
be given by \refmark{\fayet}
$$
{\cal L} = - {1 \over F} j^{\alpha \mu} \partial_{\mu} G_{\alpha}
  ~ + ~ h.c. 
\eqn\goldcoupling
$$
where $j^{\alpha \mu}$ is the supercurrent
and $G_{\alpha}$ is the spin ${1 \over 2}$ longitudinal Goldstino
component of the gravitino.
The decay to the Goldstino component is then suppressed only
by $F$ rather than $M_p$. 
In the case that the NLSP is mostly bino, $\tilde{B}$,
the coupling \goldcoupling\ 
leads to a transition magnetic dipole moment between
the NLSP and gravitino,
$
\cos \theta_W (m_{\tilde{B}} / 2 \sqrt{2} F) \tilde{B} 
\bar{\sigma}^{\mu} \sigma^{\nu} G ~ F_{\mu \nu} ~+h.c.
$, giving rise to a decay rate
$$
\Gamma( \tilde{B} \rightarrow G + \gamma) = 
{ \cos^2 \theta_W ~ m_{\tilde{B}}^5 \over 16 \pi F^2}
\eqn\binorate
$$
This translates to a decay length
$$
c \tau \simeq 130
\left( 100~{\rm GeV} \over m_{\tilde{B}}  \right)^5
\left(  \sqrt{F} \over 100~{\rm TeV} \right)^4 ~\mu{\rm m}
\eqn\decaylength
$$
So there is a range of $F$ and $m$ for which the decay
occurs within the detector, with the gravitino
carrying off missing energy.
For $m_{\tilde{B}} > m_Z^0$ there is also a non-negligible branching
fraction $\tilde{B} \rightarrow G + Z^0$
(Br$(\tilde{B} \rightarrow G + Z^0) \rightarrow \sin^2 \theta_W$
for $m_{\tilde{B}} \gg m_Z$).
In the case that the NLSP is a right handed slepton 
it can decay by
$\tilde{l}_R \rightarrow G + l_R$ with a decay
length similar to \decaylength.
If the NLSP is mostly Higgsino, 
it can decay by
$\tilde{H}^0 \rightarrow G + h^0$ if $m_{h^0} < m_{\tilde{H}}$,
where $h^0$ is the lightest Higgs boson. 
For $m_{h^0} > m_{\tilde{H}}$ decay 
$\tilde{H}^0 \rightarrow G + b \bar{b}$ is possible;
however for reasonable values of the parameters the
NLSP decays predominantly to $G+ \gamma$ 
through its gaugino components. 

Decay of the lightest standard model supersymmetric particle
to its partner plus the gravitino
within the detector gives signatures which are distinct from 
the conventional MSSM. 
Let us focus on the discovery modes at $e^+e^-$ and
hadron colliders.
Consider first the case in which the NLSP is mostly bino. 
At $e^+e^-$ colliders 
$e^+e^- \rightarrow \tilde{B} \tilde{B} \rightarrow 
\gamma \gamma + \Emiss$ is 
dominated by $t$- and $u$-channel right handed selectron
exchange. 
The production cross section for this process can be significant. 
For example,
with $\sqrt{s} = 2.2 ~m_{\tilde{B}}$, and assuming the spectrum
resulting from the simple model with $N=1$ 
given in the previous section, 
$\sigma(e^+e^- \rightarrow \tilde{B} \tilde{B} ) \simeq .87~ R$
where $R = 4 \pi \alpha^2 / 3s$ is the $e^+e^- \rightarrow \mu^+\mu^-$
cross section. 
In many models, since the bino and slepton masses
are related, the total cross section is related to the bino mass. 
This process should show significant polarization dependence
since $\tilde{e}_R$ is lighter than $\tilde{e}_L$, and 
the hypercharge of $\tilde{e}_R$ is twice that of 
$\tilde{e}_L$.
For the parameters given above 
$\sigma(e^+e^-_L \rightarrow \tilde{B} \tilde{B} ) / 
 \sigma(e^+e^-_R \rightarrow \tilde{B} \tilde{B} )
  \simeq .01$.
The bino decay is isotropic in the rest frame, implying that  
the photons have a flat energy distribution in the lab frame. 
Cuts on the $\gamma \gamma$ invariant mass can easily eliminate
the background from $e^+e^- \rightarrow \gamma \gamma Z^0$
with $Z^0 \rightarrow \nu \bar{\nu}$.


The signature $\gamma \gamma + \Emiss$ can also arise in
the conventional MSSM 
in some regions of parameter space
if the LSP is mostly Higgsino. 
In this case the NLSP 
is not much heavier than the LSP, is also mostly
Higgsino, and has a significant branching ratio
$\tilde{H}_2 \rightarrow
\tilde{H}_1 + \gamma$.
$e^+e^- \rightarrow \tilde{H}^0_2 \tilde{H}^0_2$ then gives rise
to this mode. 
In the gauge mediated case however, 
since $\Emiss$ is carried by the essentially massless
gravitinos, the photon energy is bounded by 
${1 \over 4} \sqrt{s} (1- \beta) \leq E_{\gamma} \leq
{1 \over 4} \sqrt{s} (1 +\beta)$,
where $\beta = \sqrt{1-4m^2_{\tilde{B}}/s}$ is the bino
velocity. 
In the conventional case since $\Emiss$ is carried
by the massive LSP the photon energy end points are 
smaller by a factor
$(1-m_{\tilde{H}_1^0}^2/m_{\tilde{H}_2^0}^2)$, where $\beta$ 
in this case 
is the $\tilde{H}_2^0$ velocity.  
This allows the decay to a gravitino to be distinguished 
from decay to the LSP in the conventional MSSM. 
In addition, in this region of parameter space the lightest
chargino is just slightly heavier, is also mostly Higgsino, and
decays predominantly by 
$\tilde{H}^{\pm} \rightarrow \tilde{H}^0 W^{\pm*}$.
In the MSSM the additional signatures 
$e^+e^- \rightarrow \tilde{H}^+ \tilde{H}^- \rightarrow
4j + \Emiss$, $jjl + \Emiss$, and $l^+ {l^{\prime}}^- + \Emiss$
are likely to be accessible at comparable $\sqrt{s}$.
This is in contrast to the gauge mediated case with 
a mostly bino NLSP.

As discussed in the previous section, with a 
weakly coupled messenger sector giving an NLSP
which is mostly bino, it is likely that the right
handed sleptons are not too much heavier than the NLSP.
In this case, in addition to bino pair production, slepton
pair production may be kinematically accessible.
Cascade decay through the bino then gives rise to 
$e^+e^- \rightarrow \tilde{l}^+_R 
\tilde{l}^-_R \rightarrow l^+ l^- \gamma \gamma + \Emiss$.

If the NLSP is a right handed slepton,  
the discovery mode is 
$e^+ e^- \rightarrow \tilde{l}^+_R \tilde{l}^-_R
\rightarrow l^+ l^- + \Emiss$.
As for the decay to photons, 
the leptons have a flat energy distribution,
with end points determined by $\sqrt{s}$ and $m_{\tilde{l}_R}$.
The final states with $e$, $\mu$, and $\tau$, should appear 
with very nearly equal $m_{\tilde{l}_R}$.
Cuts on $\Emiss$ can easily eliminate the 
background $e^+ e^- \rightarrow Z^0 l^+ l^-$ with 
$Z^0 \rightarrow \nu \bar{\nu}$.
This signature can also arise in the conventional MSSM
where the missing energy is carried by the massive LSP.
However, the lepton energy endpoints again distinguish
this from an essentially massless gravitino.
It is interesting to note that if $\sqrt{F}$ is much larger
than a few 1000 TeV the decay of $\tilde{l}_R$  
takes place well outside the detector.
The signature for supersymmetry is then massive charged particles,
rather than the traditional missing energy.


If the NLSP is mostly Higgsino, and $m_{\tilde{H}} > m_{h^0}$,
the discovery mode is 
$e^+e^- \rightarrow \tilde{H}^0 \tilde{H}^0 \rightarrow 
4b + \Emiss$, with
of course two pairs of $b$ jets reconstructing the Higgs mass. 
In this part of parameter space
the next heaviest neutralino and lightest chargino are
mostly Higgsino, not much heavier than $\tilde{H}^0$,
and have the same decay modes to $\tilde{H}^0$ as in the MSSM.
The signatures 
$e^+e^- \rightarrow \tilde{H}^+ \tilde{H}^- \rightarrow
4b4j+\Emiss$, $4bjjl + \Emiss$, and $4bl^+{l^{\prime}}^- + \Emiss$
should therefore also be accessible at comparable $\sqrt{s}$,
with the additional jets and leptons fairly soft. 

\REF\park{S. Park representing the CDF Collaboration,
``Search for New Phenomena in CDF,'' in {\it 10th Topical
Workshop on Proton-Antiproton Collider Physics},
edited by R. Raha and J. Yoh, AIP Press, New York, 1995.}

\REF\cdfreport{CDF Collaboration, report CDF-3456.}

\REF\paige{H. Baer, C.-h Chen, F. Paige and X. Tata,
Phys. Rev. {\bf D49} (1994) 3283.}

The discovery modes at hadron colliders
can be somewhat different than for $e^+e^-$ colliders.
If the NLSP is very nearly purely bino, 
$p \bar{p} \rightarrow \tilde{B} \tilde{B} \rightarrow
\gamma \gamma ~~+ \Emissperp$ proceeds predominantly through
$t$- and $u$- channel squark exchange, and is therefore
highly suppressed because of the large squark masses.
However, sleptons can be pair 
produced by the Drell-Yan process.
Cascade decay through the bino then leads to  
$p \bar{p} \rightarrow \tilde{l}^+_R \tilde{l}^-_R \rightarrow 
l^+ l^- \gamma \gamma + \Emissperp$.
One such spectacular $ee \gamma \gamma$ 
event has in fact been observed
at the Tevatron by the CDF collaboration 
(event 257646 in run 68739) \refmark{\park}.
The obvious background from $p \bar{p} \rightarrow WW \gamma \gamma$
has a very small production rate, and
would give rise to other decays modes which are not 
observed \refmark{\cdfreport}.
In contrast, the production cross section for
$p \bar{p} \rightarrow \tilde{l}^+_R \tilde{l}^-_R$
with $m_{\tilde{l}_R} \simeq 95$ GeV is roughly $10^{-2}$ pb
\refmark{\paige}.
With $\sim$100 pb$^{-1}$ of integrated luminosity,
the single observed event could be consistent with 
right handed slepton pair production. 
The kinematics of this event favor a fairly light bino,
implying that $e^+e^- \rightarrow \tilde{B} \tilde{B} \rightarrow
\gamma \gamma + \Emiss$ is likely to be observed at LEPII.

\REF\trilepton{H. Baer, K. Hagiwara, and X. Tata, Phys. Rev. D 
{\bf 35} (1987) 1598;
P. Nath and R. Arnowitt, Mod. Phys. Lett. A {\bf 2} (1987) 331.}

If the sleptons are much heavier than the gauginos, 
and the NLSP is mostly bino, 
pair production of winos becomes 
the dominant production mechanism, 
$p \bar{p} \rightarrow W^* \rightarrow \tilde{W}^{\pm} \tilde{W}^0$.
The dominant wino decay modes are 
$\tilde{W}^{\pm} \rightarrow \tilde{B} W^{\pm*}$
and $\tilde{W}^{0} \rightarrow \tilde{B} Z^{0*}$ through 
mixing with the Higgsino states, and 
$\tilde{W}^{\pm} \rightarrow \tilde{B} l \nu$ and 
$\tilde{W}^{0} \rightarrow \tilde{B} l^+l^-$ through off shell sleptons. 
Cascade decays through the bino then lead to the signatures
$p \bar{p} \rightarrow \tilde{W}^{\pm} \tilde{W}^0 \rightarrow
4j \gamma \gamma + \EmissT$, $jjl \gamma \gamma + \EmissT$, and
$l^+l^- l^{\prime} \gamma \gamma + \EmissT$.
The last one is similar to the standard 
tri-lepton signature of chargino pair production \refmark{\trilepton}. 
Here the additional hard photons significantly reduce the
background. 
If the NLSP is mostly Higgsino or a right handed slepton, the 
signatures at hadron colliders are similar to those at $e^+e^-$.

By far the most dramatic signature of low energy 
supersymmetry breaking 
is the possibility of measuring directly the decay of the NLSP
to its partner plus the gravitino.
If the NLSP is a neutralino this appears as a displaced
vertex, while for a slepton NLSP it appears as a kink
in a charged particle track. 
Measurement of the decay
distribution would allow a direct determination
of the supersymmetry breaking scale. 
For the decay of right handed sleptons 
to leptons, 
or the decay of Higgsinos to the lightest Higgs boson,
tracking of the resulting 
charged particles in a silicon vertex detector
and central tracking region would allow 
measurements of $c \tau$ between roughly
100 $\mu$m -- 10 m.
In the case of decay to a photon, the tracking ability
for the displaced vertex is generally not good. 
However if such a signal were established experimentally,
detectors could be optimized to convert
photons within the tracking region. 
So depending on the specific decay modes of the NLSP,
displaced vertices for 
$\sqrt{F}$ between roughly 100 -- 1000's of TeV 
could be accessible to collider experiments.

\REF\raxion{A. Nelson and N. Seiberg, Nucl. Phys. B {\bf 416}
(1994) 46; J. Bagger, E. Poppitz, and L. Randall,
Nucl. Phys. B {\bf 426} (1994) 3.} 

\REF\raffelt{G. Raffelt, Phys. Rept. {\bf 198} (1990) 1.}

This range of experimentally accessible $\sqrt{F}$ is in 
fact consistent with astrophysical and cosmological considerations. 
Unless there is an inflation with low reheat temperature,
avoiding overclosure of the universe from relic gravitinos requires
$\sqrt{F} \lsim 2 \times 10^3$ TeV.
In many theories a potentially
dangerous $R$-axion arises in the supersymmetry breaking
sector \refmark{\raxion}.
For $\sqrt{F}$ above a few TeV, 
$R$-violating interactions suppressed by a single
power of the Planck scale make the $R$-axion
too heavy to be produced during helium ignition 
in red giants \refmark{\raffelt}.
In addition, it is either trapped or too heavy to deplete
the neutrino pulse from SN1987A.
Finally, for weakly coupled models with a single additional scale,
such as the simple example in the previous section with 
$F_S \sim F$, electroweak scale superpartners are obtained for 
$\sqrt{F} \sim 100$ TeV.

\REF\cryo{P. Michelson, D. Osheroff, {\it private communication};
J. Price, in {\it Proceedings of the International Symposium
on Experimental Gravitational Physics}, eds. P. Michelson,
H. En-ke, G. Pizzella (World Scientific, Singapore, 1987)
p. 436.}

\REF\beams{M. Kasevitch, {\it private communication}.}

A final possible consequence of these theories is that 
scalar moduli with Planck suppressed couplings to matter
obtain masses of order or smaller than the gravitino
mass as the result of supersymmetry breaking. 
These fields can mediate coherent forces in the sub-millimeter
range, which has not been explored experimentally. 
New techniques employing small cryogenic mechanical oscillators
\refmark{\cryo}
or atomic beams \refmark{\beams} 
may allow the detection of such short range gravitational
strength forces.

Low energy 
gauge-mediated supersymmetry breaking clearly
makes distinct and dramatic predictions for future
experiments. The new particle
spectrum is predicted 
in terms of a small number of parameters.
For a quite plausible
range of these parameters, it predicts signatures
distinctly different than those of the conventional
MSSM. 
Most dramatic of these is the possibility of measuring 
displaced vertices or kinks in charged particle tracks
from decays to the gravitino.

We would like to thank G. Anderson, R. Barbieri, G. Giudice, 
H. Haber, L. Hall, M. Peskin, A. Pomarol, 
and J. Wells for valuable discussions.

\refout

\bye